\documentclass[aps,prb,twocolumn,showpacs,superscriptaddress]{revtex4}
\usepackage{graphicx}

\begin{document}


\title{Magnetic phase transition in V$_2$O$_3$ nanocrystals}
     \author{V.~A.~Blagojevic}    
     \affiliation{Department of Chemistry, Columbia University, New York, New York 10027, USA}
     \author{J.~P.~Carlo}
     \altaffiliation[author to whom correspondence should be addressed:  E-mail jeremy.carlo@nrc-cnrc.gc.ca]{}
     \affiliation{Department of Physics, Columbia University, New York, New York 10027, USA}
     \affiliation{Canadian Neutron Beam Centre, National Research Council, Chalk River, Ontario, K0J 1J0, Canada}
     \author{L.~E.~Brus}
     \author{M.~L.~Steigerwald}
     \affiliation{Department of Chemistry, Columbia University, New York, New York 10027, USA}
     \author{Y.~J.~Uemura}
     \affiliation{Department of Physics, Columbia University, New York, New York 10027, USA}
     \author{S.~J.~L.~Billinge}
     \affiliation{Department of Applied Physics and Applied Mathematics, Columbia University, New York, New York 10027, USA}
     \affiliation{Condensed Matter Physics and Materials Science Department, Brookhaven National Laboratory, Upton, New York 11973, USA}
     \author{W.~Zhou}
     \affiliation{Department of Applied Physics and Applied Mathematics, Columbia University, New York, New York 10027, USA}
     \author{P.~W.~Stephens}
     \affiliation{Department of Physics, State University of New York at Stony Brook, Stony Brook, New York 11794, USA}
     \author{A.~A.~Aczel}
     \affiliation{Department of Physics and Astronomy, McMaster University, Hamilton, Ontario, L8S 4M1, Canada}
     \author{G.~M.~Luke}
     \affiliation{Department of Physics and Astronomy, McMaster University, Hamilton, Ontario, L8S 4M1, Canada}

     \date{\today}
    \begin{abstract}
{V$_2$O$_3$ nanocrystals can be synthesized through hydrothermal 
reduction of VO(OH)$_2$ using hydrazine as a reducing agent. 
Addition of different ligands to the reaction produces nanoparticles, 
nanorods and nanoplatelets of different sizes. Small nanoparticles
 synthesized in this manner show suppression of the magnetic 
phase transition to lower temperatures. Using muon spin relaxation 
spectroscopy and synchrotron x-ray diffraction, we have determined
 that the volume fraction of the high-temperature phase,
 characterized by a rhombohedral structure and paramagnetism, 
gradually declines with decreasing temperature, in contrast to
 the sharp transition observed in bulk V$_2$O$_3$.}
\\
   \end{abstract}
\pacs{
}
\maketitle

\section{Introduction}

	The unique characteristics of the transition metal oxides
 make them an extraordinarily diverse class of materials, 
with properties covering almost all aspects of materials
 science and solid state physics.  The chemical and physical 
properties of the Group V oxides $-$ vanadium, niobium, 
and tantalum $-$ show promise for applications in catalysis,
 electrochemistry and electrochromical device technology.  
V$_2$O$_3$ has been the subject of many investigations 
due to a remarkable first-order metal-insulator transition at
 about 150-160K,\cite{frenkel,mcwhan} from the low-temperature 
monoclinic\cite{dernier} into the rhombohedral phase, with an
 accompanying seven order-of-magnitude increase in conductivity
and a shift from antiferromagnetic\cite{rathenau} to 
paramagnetic\cite{carr,arnold,carter1,greenwood} behavior.  
The low-temperature phase also exhibits an increase in 
volume of about 1.6\%.  The addition of a few percent of 
Cr\cite{kokabi}, Al\cite{joshi}, Re\cite{suzuki} or
 Mo\cite{tenailleau} shifts the conductivity increase to higher 
temperatures 
$-$ between 170 and 470K $-$ depending on doping fraction. 
 These properties make it possible to build temperature 
sensors and current regulators\cite{andrich,perkins} using these
 materials.  Furthermore, V$_2$O$_3$ powders have
 been used in conductive polymer composites\cite{pan} and
 in catalysis.\cite{vader} The potential uses of V$_2$O$_3$
 in the nanocrystalline form will require high chemical and 
structural purity and well-defined morphology, and careful 
characterization of the nanoparticle properties.

There have been a few successful syntheses reported of 
V$_2$O$_3$ nanocrystals.\cite{toshiyuki,zheng,su,pinna,zhang}  
Zheng \textit{et al.} obtained spherical V$_2$O$_3$ 
nanoparticles through reductive pyrolysis of ammonium oxo-vanadium(IV) 
carbonate hydroxide, 
(NH$_4$)$_2$[(VO)$_6$(CO$_3$)$_4$(OH)$_9$]$\cdot$10H$_2$O,
 in a hydrogen flow at 
650$^{\circ}$C.\cite{zheng}  Su \textit{et al.} made V$_2$O$_3$
 nanoparticles by reducing V$_2$O$_5$ crystals in high
 vacuum (10$^{-7}$ Torr) at 600$^{\circ}$C.\cite{su}  Pinna 
\textit{et al.} discovered a non-aqueous route to V$_2$O$_3$ 
nanocrystals through solvothermal reaction of vanadium(V)
 tri-isopropoxide with benzyl alcohol.\cite{pinna}  Benzyl alcohol 
plays the roles of both reducing agent and surfactant and 
the reaction is performed at 200$^{\circ}$C under nitrogen.  
Zhang \textit{et al.} synthesized V$_2$O$_3$ nanopowders
 using thermal decomposition of vanadium(IV) oxalate or mixture 
of vanadyl hydroxide and ammonium chloride in argon atmosphere
 at 500$^{\circ}$C.\cite{zhang}  Although there have been
 no successful syntheses of V$_2$O$_3$ nanorods, there have
 been a few reported syntheses of VO$_2$ nanorods, which 
we have used as a starting point.\cite{gui} 

	We report a novel approach employing hydrothermal 
reaction in water solution in air to produce V$_2$O$_3$ nanorods.  
We use different ligands to change nanocrystal shape in order 
to obtain particles, rods and platelets of V$_2$O$_3$.  
One particular improvement over the previously reported syntheses
 is that the reaction is not air-sensitive. We investigated 
the magnetic properties of small V$_2$O$_3$ nanoparticles and
 found them to behave differently from the bulk material.

\section{Synthesis}

The synthesis follows a simple route: the vanadyl hydroxide precursor 
is heated with hydrazine in an autoclave at 
230$^{\circ}$C, in water.  The reaction takes 2-14 days, depending
 on the ionic strength of the solution and the ligand used.  We were 
able to obtain different shapes and sizes of V$_2$O$_3$ nanocrystals
 through change of the ligand as shown in TEM images (Fig.~1).  
Nanorods 40-60 nm in diameter and up to 1 $\mu$m in length were
 produced using succinic acid (Fig.~1a).  Electron diffraction 
measurements of a single nanorod reveal no variation along
 the length of the rod.  Judging from the TEM images, nanorods 
account for about 10\% of the sample, the rest being nanoparticles.  
Nanorods produced using ethylene diamine as ligand have diameter
 of 40-60~nm and length of 300-500 nm (Fig.~1b).  Those produced 
using 2-propanol (Fig.~1d) as a ligand were 70-90~nm in diameter and 
around 700~nm long.  These ligands are more selective than succinic 
acid, with nanorods comprising the majority of the sample: the 
approximate yield of nanorods is around 70\% for the reaction 
using 2-propanol and about 85\% for reaction using ethylene 
diamine as the ligand.  Nanoplatelets of 80-200~nm diameter 
were obtained using ammonium citrate as ligand (Fig.~1c). Finally, 
nanoparticles (Fig.~2) were produced using 2-propanol as ligand,
 in a reaction where the precursor is washed to reduce the ionic strength
 of the solution.  The particle size distribution is broad $-$ from
 less than 10~nm to around 50~nm in diameter. 

Thermogravimetric analysis shows a gradual mass loss of about 2\%
 when sample is heated to 860$^{\circ}$C, indicating that the phase
 is stable and there is no water of crystallization.  Elemental 
analysis of the sample reveals small percentages of carbon and 
hydrogen and trace amounts of nitrogen and chlorine in the sample.
  The molar ratios of elements in the sample were consistent with 
stoichiometric V$_2$O$_3$, while the remaining carbon and hydrogen
 can almost completely be attributed to leftover ligand – succinic
 acid (C$_4$H$_6$O$_4$ $-$ 0.009 molecules per one vanadium), 
with the remaining hydrogen probably in the form of ammonium. 
 The exact origin of the excess oxygen (0.13 atoms for each vanadium
 atom) cannot be determined with any degree of certainty.

\begin{figure}
\includegraphics[width=0.46\textwidth]{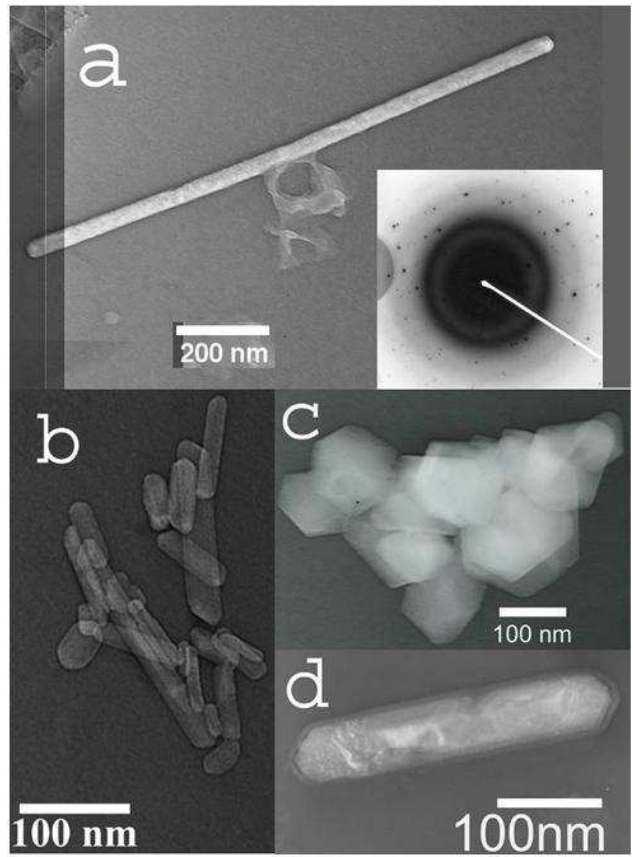}
\caption{\label{}TEM images of V$_2$O$_3$ nanocrystals synthesized
 using respective ligands: a) nanorod (succinic acid); b) nanorods
 (ethylene diamine); c) nanoplatelets (ammonium citrate); 
d) nanorod (2-propanol).}
\end{figure}

\begin{figure}
\includegraphics[width=0.46\textwidth]{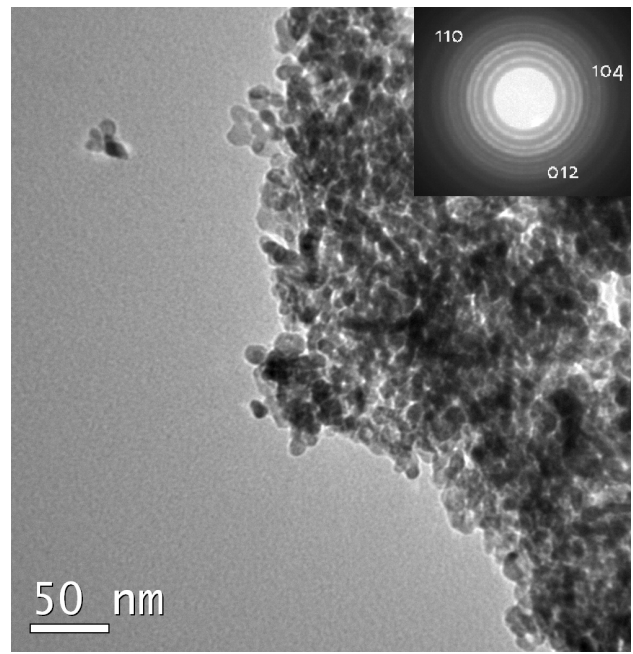}
\caption{\label{}TEM image of V$_2$O$_3$ nanoparticles 
synthesized using 2-propanol as the ligand.}
\end{figure}

\section{Results}

\subsection{Magnetic characterization}

  The temperature dependence of the magnetic susceptibility
 of V$_2$O$_3$ nanoparticles, measured using a SQUID 
magnetometer with applied fields of 100G, 200G and 1T
 (Fig.~3), show that there is no large jump in susceptibility 
near 150K, but a more gradual rise in susceptibility at lower
 temperatures. This is very different from the behavior of
 the bulk V$_2$O$_3$ crystal, which exhibits a sharp rise
 in magnetic susceptibility around 150K as the sample shifts
 from low-temperature antiferromagnetic ordering to the
 high-temperature paramagnetic state. In
external fields of 100~G and 200~G, we see the apparent
 signature of a sudden change at T~=~80K.  This, however, 
can be explained as spin freezing of a small impurity phase or 
the onset of a slight ferrimagnetic polarization
of primarily antiferromagnetic spins, as even the largest values of
 susceptibility seen at low temperatures correspond to a moment
 of approximately only 2$ \times $10$^{-4}$ $\mu_B$ per V site.  
In addition, while the high-temperature state in bulk V$_2$O$_3$
 is characterized by a relatively constant Pauli-like susceptibility,
 the nanopowder sample exhibits distinctly Curie-Weiss-like behaviour 
at high temperatures.  In order to further investigate this behavior, 
we used muon spin relaxation spectroscopy.

Muon spin relaxation ($\mu$SR)\cite{schenck,muonsci} is an 
experimental technique based on the implantation of spin-polarized 
muons in matter and the influence of the atomic, molecular or 
crystalline surroundings on the evolution of their spins. It is a
 point-like magnetic probe in real space, similar to NMR and ESR, 
and complementary to the frequency space probed by scattering 
measurements. One key difference between $\mu$SR and other 
related probes such as NMR and ESR is the time window in detecting 
dynamic phenomena: $\mu$SR is sensitive to time scales over the range
 of  10$^{-11}$ to 10$^{-5}$~s. As an added advantage, $\mu$SR
 can be performed in the absence of an external magnetic field.

In a magnetic field, the muon spin precesses at a frequency of about 
13.5~kHz~G$^{-1}$; the magnetic moment of the muon is about three
 times that of the proton, making it a very sensitive magnetic probe.
 The positive muons used in the measurement, produced by the 
parity-violating weak decay of pions at rest, exhibit almost 100\%
 initial spin polarization, and decay into a positron and two 
neutrinos with a mean lifetime of 2.2$\mu$s. The positron is 
emitted preferentially along the direction of the instantaneous 
muon spin moment at the time of decay, with a typical overall 
20-30\% asymmetry between the directions parallel and antiparallel
 to the muon spin at time of decay. The time-dependent asymmetry
 of detected decay positrons correlates directly to the time 
evolution of the spins of the implanted muons, and thus to the
 internal local field distribution in the sample. 

Temperature-dependent $\mu$SR measurements\cite{uemura,savici}
 were performed at the M15 beamline at TRIUMF in a transverse field
 of 50~G in the temperature range of 3-180K (Fig.~4) using the 
\textit{HiTime} spectrometer. The long-lived
 oscillating component represents muons in paramagnetic environments,
 while precessing muon spins in the antiferromagnetic environment 
quickly dephase due to local fields, resulting in rapid relaxation of
 the initial muon spin asymmetry. The data indicates that the onset
 of magnetic ordering occurs at different temperatures in different
 parts of the sample (presumably corresponding to different
 particle sizes).   This is in stark comparison to the rapid and hysteretic
transition previously
observed in $\mu$SR measurements of bulk V$_2$O$_3$. \cite{uemura84}

\begin{figure}
\includegraphics[width=0.46\textwidth]{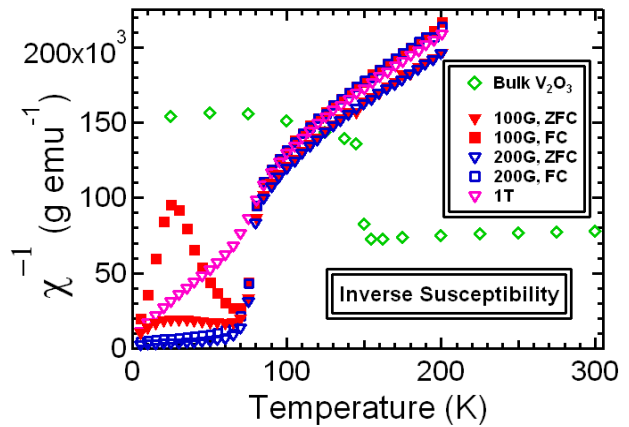}
\caption{\label{}Magnetic susceptibility of V$_2$O$_3$ nanoparticles
 as a function of temperature, in applied fields of 100G (red solid 
symbols), 200G (blue open symbols) and 1T (violet open symbols), 
with susceptibility data of McWhan \textit{et al.}\cite{mcwhan} 
for the bulk sample (green diamonds) 
included for reference.  Zero-field cooled (ZFC) and field-cooled 
(FC) measurements are denoted by inverted triangles and squares, 
respectively.}
\end{figure}

\begin{figure}
\includegraphics[width=0.46\textwidth]{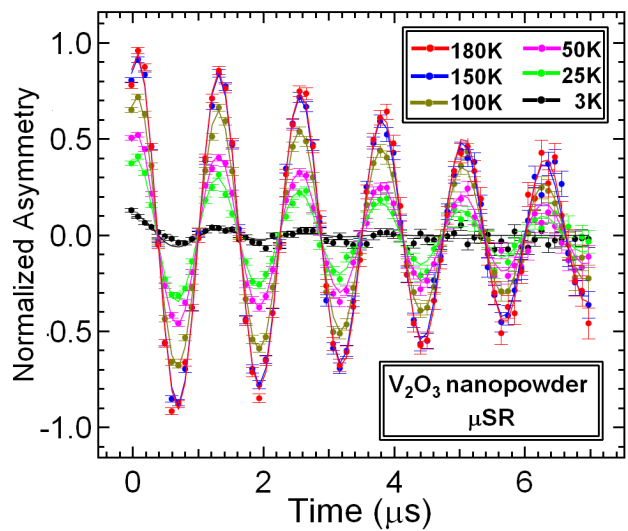}
\caption{\label{}Muon spin relaxation measurements performed
 in a weak transverse magnetic field (wTF) of nominally 50G, as 
a function of temperature from 3K to 180K.}
\end{figure}

\subsection{Structural Characterization}

Bulk V$_2$O$_3$ exhibits concurrent magnetic, structural and 
metal-insulator phase transitions.  To attempt to correlate the 
observed magnetic properties of V$_2$O$_3$ nanopowder 
with any structural changes, we performed synchrotron x-ray 
diffraction measurements on our sample, as well as on commercially
 prepared bulk powder.  These measurements were conducted at
 beamline X16C at the National Synchrotron Light Source (NSLS) at
 Brookhaven National Laboratory (BNL), with the samples packed in
 quartz capillaries and measured using x-rays of wavelength 0.698955~$\AA$. 
 Complete datasets from 10$^{\circ}$ to 40$^{\circ}$ were collected
 for the bulk sample as a function of temperature, and complete data 
sets at 20K and 300K, as well as shorter scans at intermediate 
temperatures, were taken for the nanopowder sample.

The bulk data were analysed using the \textit{FullProf} Rietveld
 refinement program,\cite{rodriguez} and the data were fit to a
 rhombohedral high-temperature phase and a monoclinic low-temperature
 phase consistent with earlier published results. We confirm that, 
as in earlier published results, \cite{frenkel,mcwhan} the transition occurs quite rapidly,
 and is associated with coexistence of the high- and low-temperature
 phases over a narrow range of temperatures, with the transition
 temperature exhibiting significant hysteresis, as expected for a 
strongly first-order phase transition.

In the nanopowder sample, broadening of Bragg peaks 
and a high background made Rietveld refinement impossible
 and necessitated much longer integration times for data collection.
Measurements of the empty quartz capillary showed that this large
 background phase is in fact intrinsic to the sample, although it is
 temperature-independent, and its origin remains unknown.

As a result of these limitations, we chose to focus on the region
 around 24$^\circ$ corresponding to the bulk [116] peak. 
 In the bulk sample, this peak exhibits a three-fold splitting as the
 sample transitions from the rhombohedral to the monoclinic phase. 
Although the [116] peak in the nanopowder sample overlaps with 
a broad background peak, this background is temperature-independent
 and may be subtracted from the data after being fitted to a 
temperature-independent Gaussian.  The sample signal was fit to a
 sum of a single Gaussian representing the high-temperature phase, 
and a sum of three Gaussians representing the low-temperature phase. 
 Our results are shown in Fig.~5, revealing the gradual dropoff of
 the high-temperature peak intensity with decreasing temperature,
and a corresponding rise in the more diffuse scattering representing
 the low temperature phase.  While limitations of the data do not 
allow us to quantitatively analyze the structure of the 
low-temperature phase, we can determine that coexistence of
the high- and low-temperature phases does occur over a wide 
temperature range similar to that where the high- and low-temperature
 magnetic phases were seen to coexist in the  $\mu$SR measurements. 

This motivated us to compare the relative phase fractions of high and
low-temperature phases obtained by the two techniques.  In Fig.~6 we 
show in red  the temperature dependence of the oscillating asymmetry
 component of muon spin relaxation, giving a relative measure of
 volume fraction of the paramagnetic region, and in blue the
 temperature dependence of the integrated intensity coming from the
 high-temperature structural phase observed by x-ray diffraction.  
For comparison, the high-temperature rhombohedral phase fraction 
observed by x-ray diffraction in the bulk material are shown for both
 warming and cooling cycles.  Both methods reveal that the volume
fraction of the rhombohedral/paramagnetic phase increases 
gradually with increasing temperature and levels off after T $\sim$ 150K, 
suggesting that these signals come from the same regions of the material.

\begin{figure}
\includegraphics[width=0.46\textwidth]{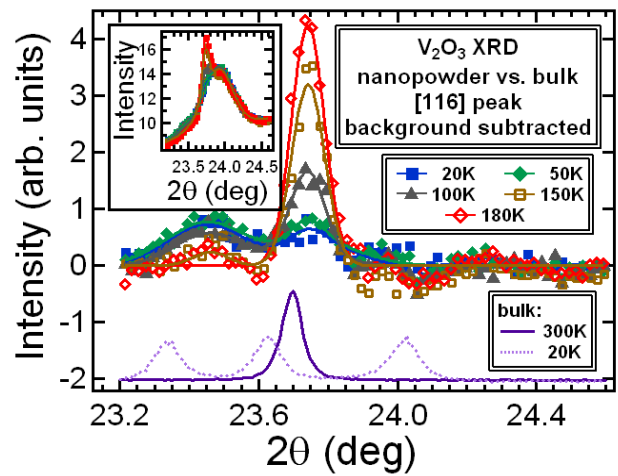}
\caption{\label{}XRD Data around the [116] peak.  The nanopowder
 data is shown with a temperature-independent background (seen 
as the broad peak in the inset) subtracted.  Comparison data for
 the bulk material is shown at the bottom (arbitrary intensity 
offset).  The high-temperature phase is represented by the 
single peak near 23.75$^\circ$, whereas the low-temperature 
phase is seen in several broad peaks, most noticeably near 
23.4$^\circ$.  A gradual crossover is seen between the two 
phases as a function of temperature.}
\end{figure}

\begin{figure}
\includegraphics[width=0.46\textwidth]{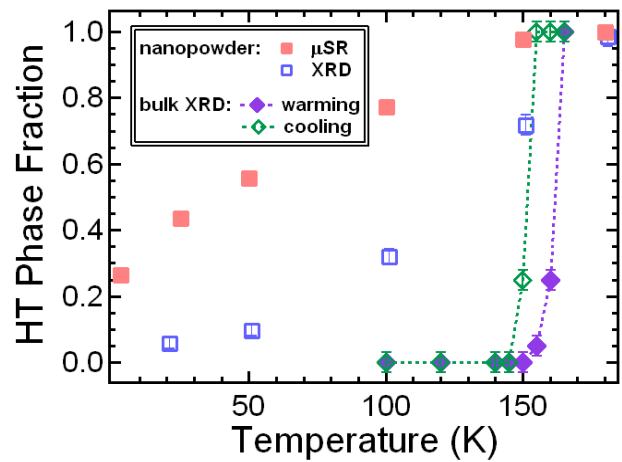}
\caption{\label{}Comparison of high-temperature phase fractions 
determined by $\mu$SR (red solid squares) and XRD (blue open 
squares). Corresponding phase fractions determined by XRD on 
bulk V$_2$O$_3$ on warming (violet filled diamonds) and cooling
 (green open diamonds) are shown for reference.}
\end{figure}

\section{Discussion}

\subsection{Synthesis}

The vanadyl hydroxide precursor obtained through different 
processes yields different results. When it is obtained by 
reducing V$_2$O$_5$ with hydrazine, the reaction yields
 V$_2$O$_3$. When the precursor is prepared by precipitation
 from VOSO$_4$ solution without a reducing agent, the
 reduction to vanadium(III) does not occur. However, when the
 same reaction is performed with the addition of hydrazine to 
the precursor before placing it in the autoclave, the result is 
V$_2$O$_3$. This suggests that excess hydrazine is acting 
as a reducing agent in the final step of the reaction, reducing 
vanadium(IV) to vanadium(III). In fact, there is evidence in the
 literature\cite{su} suggesting that if the molar ratio of 
hydrazine to vanadium is 1:1, allowing only for reduction from
 vanadium(V) to vanadium(IV), that the product of hydrothermal 
reaction is VO$_2$$\cdot x$H2O. Reactions performed at temperatures
 below 200$^{\circ}$C failed to produce V$_2$O$_3$, suggesting that 
there is a minimum temperature required for vanadium reduction under
 hydrothermal conditions. 2-propanol accelerates the reaction as 
compared to equivalent ligand-free reaction: the reaction with 
2-propanol as a ligand takes 1 day to produce 
V$_2$O$_3$ at 230$^{\circ}$C, while a ligand-free reaction 
under the same conditions takes 3 days. There are two possible 
explanations, which are not mutually exclusive: 2-propanol might 
be acting as a weak reducing agent, thus accelerating vanadium 
reduction, or the acceleration might simply come from the fact that
 the 2-propanol reaction produces much smaller particles than any
 other we have performed. Small particles might be reduced more 
quickly simply because they would expose a larger surface area to
 hydrazine in the solution. It is also interesting to note that the
 ionic strength of the solution appears to have an effect on both
 the reaction time and the morphology of the product. We can
 look at two reactions using 2-propanol as a ligand and differ only 
in the fact that, in the second one, the precursor was washed with
 water to reduce ionic strength. The first reaction takes 8 times 
longer to complete and produces a mix of nanorods and nanoparticles
 (70-90~nm in diameter). The second reaction produces only small 
nanoparticles (7-50~nm in diameter). We also observed that using 
ethylene diamine as a ligand requires significantly longer reaction
 time (14 days) than other reactions at the same conditions (7 days). 
This suggests that ethylene diamine slows down the reaction, possibly
 by stabilizing vanadium(IV).

\subsection{Characterization}

Magnetic measurements on V$_2$O$_3$ nanoparticles synthesized 
using 2-propanol as ligand show that the magnetic transition 
(from anti-ferromagnetic to paramagnetic) appears to be occurring
gradually as the temperature increases, rather than abruptly as it
 does in the bulk crystals. SQUID measurements (Fig.~3) show a
 gradual inflection in susceptibility below 150K, and no bulk-like 
sharp transition at 150-160K. Analysis of the muon spin relaxation and 
x-ray diffraction measurements (Fig.~6) show that the volume of
 the paramagnetic/rhombohedral region increases steadily with
 increasing temperature,suggesting a particle size dependence on
transition temperature. By the time the temperature reaches 
150K the curve of the asymmetry amplitude dependence is 
almost leveled off. The asymmetry value at 180K
 is about 0.16, indicating that all of the nanoparticles are
 in the paramagnetic state.  The x-ray diffraction results
 largely follow this trend, indicating a gradual decrease of the 
high-temperature phase fraction with decreasing temperature.  
Characterization of the crystalline phases is impossible, however, 
due to the limitations of the data, primarily from a large but
 temperature-independent signal arising from the sample itself.  
We are thus limited to a semi-quantitative estimate of the phase
fractions, which display the same trend as the magnetic data.

Magnetic ordering of the monoclinic phase of bulk V$_2$O$_3$ has been revealed
by polarized neutron scattering experiments to be
 ferromagnetic in planes perpendicular to the [010] direction, 
while adjacent planes are antiferromagnetically coupled\cite{moon,heidemann}. 
This results in zero overall magnetic moment, 
event though individual layers have a relatively high moment of 
1.2$\mu_B$ per vanadium atom. Experiments\cite{carter2} have 
shown that the magnetic transition can be shifted to a temperature
 below 150K by application of pressure. At a pressure of about 
15 kbar the magnetic transition decouples from the metal-insulator 
transition, and by about 26~kbar, both phase transitions are 
completely suppressed. 

\subsection{Possible Size Dependence of Transition Temperature}

It is apparent from the XRD data that the high-temperature [116] peak in the nanopowder
sample is shifted to an angle about 0.2\% larger than is seen in the bulk material, indicating
that the lattice parameters in the nanopowder sample are correspondingly smaller than in the bulk.  
Furthermore, the [116] peak in the bulk material exhibits significant broadening relative to its
counterpart in the bulk.  Bragg peak widths can be ascribed to several causes,
including instrumental resolution, finite particle size, variation of lattice parameters throughout
different parts of the specimen, and microstrain effects.  The width of the broadened nanopowder peak,
well approximated by a Gaussian, may be parametrized as $\sigma_{NP}^2 = \sigma_{int}^2 + 
\sigma_{exc}^2$, where $\sigma_{int}$ is the intrinsic width approximated by the width of the peak
 in the bulk material, and $\sigma_{exc}$ defined as the ``excess width'' due to the remaining 
factors.

In our current data set, the FWHM (full width at half maximum) in 2$\Theta$ of the bulk peak is
0.152$^\circ$, while that of the nanopowder sample is 0.222$^\circ$; this leaves an excess
FWHM due to broadening of 0.162$^\circ$.  If we ascribe the entirety of this broadening
to the finite size of the nanoparticles, we find by application of the Debye-Scherrer formula\cite{patterson}
$L = K\lambda/\beta $cos$\Theta$ (where $L$ represents the particle size, $\lambda$ the wavelength, 
$\Theta$ the Bragg angle and $\beta$ the peak FWHM in radians) 
that the observed width corresponds to an average particle size of 23 nm, in the lower mid-range of the 
10$-$50 nm (based on TEM images) particle size distribution, assuming a typical Debye-Scherrer constant
value $K = 0.9$.  Unfortunately, it 
is impossible to distinguish between the varying sources of broadening in the current data set, nor is it 
possible to determine whether finite size effects alone are sufficient to explain the observed broadening.
 
One intriguing interpretation of the range of transition temperatures observed is that it arises from
the relatively wide particle size distribution in the sample. 
In particular,  since progressively smaller nanoparticles have an increasingly large ratio of surface area
to volume, it is expected that surface tension would hold smaller particles in greater compression, resulting in 
smaller lattice parameters.  If the lattice parameters are a critical determinant of the transition 
temperature, the transition temperature and its range over the sample may be determined from measurement 
of the lattice parameters.

Finger and Hazen\cite{finger} have determined the variation of the lattice
parameters of bulk V$_2$O$_3$ under applied pressure up to 50~kbar, from which we may compute
the [116] peak d-spacing as a function of pressure.  McWhan \textit{et al.}\cite{mcwhan}
have characterized the depression of the transition temperature as a function of applied pressure, 
finding that the transition temperature
drops linearly at a rate of approximately $-$5~K/kbar up to about 16~kbar, then more rapidly decreases 
toward zero at a sample-dependent critical pressure of 20$-$26~kbar.  Combining these results, one 
may estimate approximately the transition temperature as a function of [116] d-spacing.  Using this relation, 
we find that the centroid of the nanopowder [116 ] peak corresponds to an effective pressure of 14~kbar and
 an estimated transition temperature of 85 K.   If there is a spread of lattice parameter
values throughout the sample, it would also follow that this transition should be smeared out, with portions of
the sample exhibiting smaller lattice parameters (corresponding to a higher effective pressure) undergoing the
transition at lower temperatures, and portions of the sample exhibiting larger lattice parameters (corresponding
to effective pressures closer to zero) undergoing the transition at higher temperatures.
This is roughly consistent with our $\mu$SR and XRD observations 
indicating coexistence of the high- and low-temperature phases from base temperature up to over 150K,
although uncertainties about the precise distribution of particle shapes and sizes and unknown strain effects
prevent a more quantitative comparison.

Finally, since magnetic ordering in the low-temperature phase is long-range, and because of the dimensions 
of the nanoparticles, the magnetic transition caused by the anti-parallel alignment of magnetic moments in 
adjacent planes might not necessarily occur concurrently with the structural transition. Further investigation 
is required to confirm these findings.

\section{Conclusions}

We present a novel approach for synthesis of
V$_2$O$_3$ nanocrystals via hydrothermal reduction. 
Unlike any other reported syntheses, this reaction is done
 in water and under ambient atmosphere. We have obtained 
nanocrystals of varied shapes and sizes through 
choice of ligand. 

The magnetic properties of V$_2$O$_3$ nanoparticles have
 been measured with respect to temperature using susceptibility 
and muon spin relaxation, and it is found that unlike the behavior
in the bulk material, there is a gradual onset of magnetic order with 
decreasing temperature.

Additionally, the structural properties of V$_2$O$_3$ nanoparticles 
have been studied using synchrotron x-ray diffraction.  While a full 
structural refinement of the data was impossible due to
 the large background signal, clearly a 
high-temperature phase and a low-temperature phase trade off
with one another as a function of temperature.
Both the magnetic and the structural results indicate coexistence 
of high- and low-temperature phases over a broad temperature
 range, with qualitative agreement between the phase fractions 
determined by magnetic and by structural measurements.  The 
high-temperature phase is consistent with the observed bulk
 rhombohedral phase, although the low-temperature 
phase structure cannot be deduced with the available data.

These results suggest a correspondence between lattice parameters 
and transition temperature, with an intriguing explanation being that a particle
size dependence to the transition temperature exists due to compression by surface tension.
However,  the relatively wide size distribution of the nanoparticle sample precludes any 
definitive conclusions, which await further investigation.

\section{Acknowledgments}

{We gratefully acknowledge T. J. Williams, G. J. MacDougall, J. A. Rodriguez, 
J. Janik and C. R. Wiebe for help in $\mu$SR data acquisition, Emil Bozin 
and Peng Tian for help with the diffraction data analysis, I.~P.~Swainson and A.~J.~Millis 
for useful discussions, and the TRIUMF CMMS staff for invaluable technical
assistance with $\mu$SR experiments.  Work at Columbia University
was supported by the U.S. National Science Foundation via grants DMR-0502706, 
DMR-0703940 and DMR-0806846 and the MRSEC program 
(DMR-0213574),  and at McMaster University by NSERC and the Canadian 
Institute for Advanced Research.    W. Zhou was supported by NSF grant DMR-0520547.
Work in the Billinge group, and use of the National Synchrotron Light Source, Brookhaven National Laboratory, 
was supported by the U.S. Department of Energy, Office of Basic Energy Sciences, 
under Contract No. DE-AC02-98CH10886.}    Finally, we thank the NSF Partnerships for
International Research and Education (PIRE) program for valuable support.



\twocolumngrid

\end{document}